\documentclass[onecolumn,aps,showpacs,superscriptaddress,floatfix,
               preprintnumbers,amsmath,amssymb,nofootinbib]{revtex4}
\usepackage{graphicx}
\usepackage{dcolumn}
\usepackage{bm}

\begin{document}
\title{Ultrahigh-Energy Cosmic Ray Particle Spectra from Crypton Decay}

\author{John Ellis}
\affiliation{Theory Division, Physics Department, CERN, CH 1211 Geneva 
23, 
Switzerland}
\author{V.E. Mayes}
\affiliation{George P. and Cynthia W. Mitchell Institute for 
Fundamental Physics, Texas A\&M 
University,\\ College Station, TX 77843, USA}
\author{D.V. Nanopoulos}
\affiliation{George P. and Cynthia W. Mitchell Institute for 
Fundamental Physics, Texas A\&M 
University,\\ College Station, TX 77843, USA}
\affiliation{Astroparticle Physics Group, Houston
Advanced Research Center (HARC),
Mitchell Campus,   
Woodlands, TX~77381, USA; \\
Academy of Athens,
Division of Natural Sciences, 28~Panepistimiou Avenue, Athens 10679, 
Greece}

\begin{abstract}
We calculate the spectra of ultra-high-energy cosmic rays (UHECRs) in an
explicit top-down model based on the decays of metastable neutral
`crypton' states in a flipped SU(5) string model. For each of the
eight specific $10^{th}$-order superpotential operators that might
dominate crypton decays, we calculate the spectra of both protons and
photons, using a code incorporating supersymmetric evolution of the
injected spectra. For all the decay operators, the total UHECR spectra are
compatible with the available data. Also, the fractions of photons are 
compatible with all the published upper limits, but may be detectable in 
future experiments.
\end{abstract}
\maketitle

\begin{center}
CERN-PH-TH/2005-251~~~~~~MIFP-05-34~~~~~~ACT-13-05~~~~~~{\tt
astro-ph/0512303}
\end{center}

\section{Introduction}

The existence of cosmic rays with energies above the
Greisen-Zatsepin-Kuzmin (GZK)  
cutoff~\cite{Greisen:1966jv,Zatsepin:1966jv} is one of the most important
open problems in high-energy astrophysics~\cite{Nagano:2000ve,Westerhoff}.  
These ultra-high energy cosmic rays (UHECRs) may be a tantalizing hint of
novel and very powerful astrophysical accelerators, or they may be
harbingers of new microphysics via the decays of metastable supermassive
particles. {\it It is remarkable that we still do not know whether the
UHECRs originate from macrophysics or microphysics.} If there is no GZK
cutoff, as suggested by the AGASA data~\cite{AGASA}, the sources of the
UHECRs would need to be local. In this case, since local magnetic fields
are unlikely to have deflected significantly their directions of
propagation, the UHECRs would `remember' the directions of their sources.
Thus, one would expect some anisotropy in the arrival directions of the
UHECRs, associated either with discrete energetic astrophysical sources
nearby, such as BL Lac objects~\cite{Westerhoff}, or the 
distribution of (mainly galactic) superheavy dark matter. No significant
anisotropy of the UHECRs has yet been seen, but the existing experiments
have insufficient statistics to exclude one at the expected 
level~\cite{Evans}. On the
other hand, the GZK cutoff may be present in the HiRes data~\cite{HiRes},
in which case no exotic microphysics may be needed, and any astrophysical
sources would be less restricted and more difficult to trace. The first
batch of Auger~\cite{Sommers:2005vs} data are inconclusive on the possible
existence of the GZK cutoff.

Superheavy particles of the type required could have been produced
gravitationally around the end of
inflation~\cite{inflation}.  
Particularly interesting candidates for such superheavy particles are \lq
cryptons\rq, bound states of the fractionally-charged constituents that
arise generically~\cite{Schellekens:1989qb} in the hidden sectors of
models of particle physics derived from the heterotic
string~\footnote{Such states may also be a generic feature of models
constructed from intersecting D-branes.}. Cryptons arising in the hidden
sector of a heterotic string-derived flipped SU(5) model may have exactly
the right properties to play this
role~\cite{cryptons,EMN1}.  These \lq flipped
cryptons\rq \ are bound by SU(4) hidden-sector interactions, and include
four-constituent meta-stable bound states called {\it tetrons} that are
analogues of the three-constituent baryons of QCD, as well as
two-constituent meson-like states. Indeed, it was within this flipped
SU(5) model that the confinement solution to avoiding the stringent
experimental limits placed on fractional charges was first pointed out,
and this model remains the only example to have been worked through in any
detail~\cite{cryptons,EMN1}.

In general, tetrons may decay through 
$N^{th}$-order non-renormalizable
operators in the superpotential, which would yield lifetimes that are
expected to be of the order of 
\begin{equation} 
\tau \approx \frac{\alpha_{string}^{2 - N}}{m_X} \left( \frac{M_s}{m_X} 
\right)^{2(N-3)}, 
\label{lifetime} 
\end{equation}
where $m_X$ is the tetron mass and $M_S \sim 10^{18}$~GeV is the string
scale. The $\alpha$-dependent factor reflects the expected dependence 
of high-order superpotential terms on the effective gauge coupling $g$. 
The mass scale associated with these states is estimated using the
renormalization group for the SU(4) interactions to be $\Lambda 
\sim10^{12}-10^{13}$~GeV, just in the right range for their decays to 
produce the UHECRs.

The lifetimes of neutral tetrons without electric charge has been
estimated to lie in the range $\tau_0 \sim 10^{11} - 10^{17}$~years, so
that they may still be present in the Universe today, and might produce
the necessary flux of UHECRs if they are sufficiently abundant.  We have
shown in our previous work~\cite{EMN1} that the mesons and charged
tetrons - whose present-day abundances are subject to very stringent
experimental limits - would have decayed with short lifetimes early in the
Universe. In the course of studying the possible tetron lifetimes, we
identified various 10$^{th}$-order superpotential operators that might
govern neutral tetron decays. Thus, in this specific model of flipped
cryptons we are able to go beyond generic statements regarding the
injected UHECR particle spectra that may result from their decays, and
make a number of specific predictions.

This enables us to address an important experimental constraint on such
crypton models of the UHECRs. Although \lq top-down\rq \ models such as
crypton decays may appear to be natural explanations for the UHECRs (if
they exist), they generically share a potential drawback. The spectra of
UHECRs that they produce might be expected to have large photon fractions,
in possible conflict with the observation that most of the UHECR primaries
appear to be protons or nuclei. The Auger collaboration has recently set
an upper limit of $26\%$ on the photon fraction above
$10^{19}$~eV~\cite{Risse:2005hi}, and the limit may even be as low as
7-14\%~\cite{Sarkar}), while an upper limit of $50\%$ at energies above
$10^{20}$~GeV has been set by the AGASA collaboration~\cite{Risse:2005jr}.

In this paper, we first give a review of relevant aspects of the flipped
SU(5)  heterotic string model. We then analyze the specific primary
multi-body decay modes governed by the various different 10$^{th}$-order
superpotential operators found in our previous paper. We then calculate
the UHECR particle spectra that would be injected by these decays using
one of the most detailed and complete codes currently
available~\cite{Barbot:2003wv}, paying particular attention to the photon
fraction. The total UHECR spectra obtained from the various superpotential
terms do not differ greatly. On the other hand, we find that different
decay operators may give rather different photon fractions, particularly
at the highest energies. However, in every case, we find that the
calculated spectra after supersymmetric evolution and fragmentation are
compatible with the published upper limits on the photon fractions in
various energy ranges, when we include the UHECR background resulting from
a homogenous extragalactic distribution of sources and incorporate the
pile-up expected from the GZK effect. There is no need to appeal to the
cosmic radio background to absorb a significant fraction of the photons in
order to bring them below the AGASA and other limits.

\section{Generic Super-heavy Relic Decay}
The basic idea in generic \lq top-down\rq~explanations (see ~\cite{Berezinsky:1997hy,Kuzmin:1997cm}) is that the UHECR are produced via the decay of some
relic particles or topological defects left over from the inflationary epoch and which are locally clustered in the galactic halo as cold, dark matter with an over-density $n_X/n^{cos}_X\sim10^{4-5}$.  The lifetime of such relics must exceed the present age of the universe in order for them to exist today in sufficient abundance, however the lifetime must not be too large so that the decay rates produced are too small to produce the UHECR.  Furthermore, the relic mass must be at least $M_X > 10^{12}$~GeV in order to produce the UHECR energies observed.  Typically, the lifetime of a particle is expected to be inversely proportional to it's mass, $\tau \sim1/M$.  Clearly it is not easy to have a particle with both a large enough mass and a decay lifetime in the right range to produce the UHECR.  However, as pointed out in the Introduction, flipped $SU(5)$ cryptons satisfy both of these criteria, which makes them very attractive as a top-down explanation of the UHECR.  Indeed the \lq flipped\rq~crypton is probably the most natural and most physically motivated of any top-down candidate. 

There are three generic statements that can be presently made about a decaying super-heavy $X$ particle explanation for the UHECR:
\begin{enumerate}
	\item Since the super-heavy relics may accumulate locally in the galactic halo with an over-density $\sim10^{4-5}$ over the cosmological average, they may avoid the GZK cutoff.
	\item Due to the displacement of our solar system with respect to the galactic plane, there should be some anisotropy in the arrival directions of the UHECR with respect to the galactic center.  
	\item Photons tend to be the dominant component of the UHECR flux produced by the super-heavy $X$ decay.  However,
the photons may scatter off the galactic radio background, which is poorly measured, and thus may be somewhat attenuated.
\end{enumerate}
  
The injection spectrum produced by such a decaying super-heavy relic $X$-particles with number density $n_X$ and lifetime $\tau_X$ is proportional to the inclusive decay width:
\begin{equation}
\Phi^{halo}(E) = \frac{n_X}{\tau_X} \frac{1}{\Gamma_X} \frac{d\Gamma(X \rightarrow g_1 + \cdots)}{dE}.
\end{equation}
If a spherical halo of radius $R_{halo}$ and uniform number density $n_X$ is assumed, then the galactic halo contribution
to the UHECR will be given by
\begin{equation}
J^{halo} = \frac{1}{4\pi} R_{halo}\Phi^{halo}(E).
\end{equation}
In general, the $X$-particles will decay into one or more partons which hadronize into other particles $g$ of the MSSM, which carry a fraction $x$ of the maximum available momentum $M_X/2$ and a fraction $z$ of the parton momentum.  For such a decay,
the inclusive decay width can be factored as 
\begin{equation}
\left.\frac{1}{\Gamma_X} \frac{d\Gamma(X \rightarrow g_1 + \cdots)}{dE} = \sum_a \int_0^x \frac{1}{\Gamma_a} \frac{d\Gamma_a(y,\mu^2,M_X^2)}{dy}\right|_{y=x/z} D^g_a(z,\mu^2),
\end{equation}
where $D^g_a(z,\mu^2)$ is the fragmentation function (FF) for particles of type $g$ into particles of type $a$, and $\mu$ is the energy scale, most appropriately taken to be equal to the $X$ particle mass, $\mu = M_X$.  The evolution
of the fragmentation function is governed by the DGLAP equations, which may be extended to include the MSSM.  Thus, the determination of the expected UHECR flux from the super-heavy $X$ decay essentially becomes the problem of starting with a set of initial decay partons and evolving the decay cascades via the fragmentation functions to find the end decay products and energy distribution.  To evolve the fragmentation functions up to the energy of the super-heavy $X$ decay, the DGLAP equations must be solved numerically. Several groups have done such calculations for generic initial decay partons (usually into a quark-antiquark pair)~\cite{Birkel:1998nx, Berezinsky:2000up, Barbot:2002ep, Aloisio:2006yi}.  Perhaps the best such code is \texttt{SHdecay}~\cite{Barbot:2003wv}. 
This code 
calculates the fragmentation into the seven stable MSSM particles 
($p$, $\gamma$, $e$, neutralino LSP $\chi$, $\nu_e$, $\nu_{\mu}$, 
$\nu_{\tau}$) for any given initial decay parton.  In the case of flipped $SU(5)$ cryptons, we have a specific model
where the initial decay partons in the cascade are known.         

In addition to the UHECR flux produced by the super-heavy decay from $X$ particles clustered in our galactic halo,
there may be a background flux from sources outside of our galaxy, perhaps super-heavy $X$ decay in \textit{other} galactic halos or in intergalactic regions. Generally, this flux is assumed to be due to a homogenous distribution of sources and exhibits a characteristic GZK pileup due to the fact that they are produced \textit{non-locally}~\cite{Berezinsky:2005cq}.  Since the GZK attenuation is much more severe for photons than for nucleons, this background should be comprised primarily of nucleons.  Thus, in this scenario the UHECR flux observed on Earth will be the sum of this extragalactic background and the local galactic flux from decaying relics clustered in our galactic halo. Due to the extragalactic component, super-heavy $X$ particle decay may only unambiguously explain the UHECR flux for energies $E > 4\cdot10^{19}$~eV. However, we note that the extragalactic component may also be due partially to \textit{non-local} super-heavy $X$ decay, as well from astrophysical sources (\lq bottom-up\rq~production). Thus, it is  more accurate to say that a distinct signal of super-heavy $X$ decay within our galactic halo would be the existence of an excess of events above this energy.  The lack of any events above this energy would not rule out the presence of a top-down component, but it would not provide an unambiguous reason for the introduction of such a mechanism.

\section{String-Derived Flipped SU(5)}

Even before the first string models, there was strong interest in flipped
SU(5) as a possible grand unified theory, primarily because it did not
require large Higgs representations and avoided the constraints of minimal
SU(5)~\cite{GG} without the extra gauge interactions required in larger
groups such as SO(10)~\cite{Barr:1981qv,Derendinger:1983aj}. Interest in
flipped SU(5) increased within the context of string theory, since simple
string constructions could not provide the adjoint and larger Higgs
representations required by other grand unified theories, whereas they
could provide the Higgs representations required for flipped SU(5).  It
was, moreover, observed that flipped SU(5) provided a natural
`missing-partner' mechanism for splitting the electroweak-doublet and
colour-triplet fields in its five-dimensional Higgs
representations~\cite{AEHN}. Heterotic string-derived flipped SU(5)
remains among the most fully developed and realistic models yet derived
from string.

Flipped SU(5) derives its name from the flipping of the quark and 
lepton assignments relative to those in the
minimal SU(5) GUT~\cite{GG}: $u_L, u^c_{L} 
\leftrightarrow d_L, d^c_{L}$ and 
$\nu_L, \nu^c_{L} \leftrightarrow e_L, e^c_L$ for all three generations.
This requires the gauge group to become SU(5) $\times$ U(1), so as to 
accomodate the charges of the quarks and leptons, which fill up the 
$\bar{\mathbf{5}}$, $\mathbf{10}$ and $\mathbf{1}$ representations of SU(5):
\begin{equation}
     f_{\mathbf{\bar{5}}}  =  \begin{pmatrix} u^c_1\\ u^c_2\\ u^c_3\\ 
e\\ \nu_e
                     \end{pmatrix}_L; \ \ \ \
     F_{\mathbf{10}}       =  \begin{pmatrix} 
                       \begin{pmatrix} u\\ d
                       \end{pmatrix}_L
                     d^c_L & \nu^c_L
                     \end{pmatrix}; \ \ \ \
     l^c_{\mathbf{1}} = e^c_L .
   \label{FSU5Rep}
\end{equation}
The presence of a neutral component in the $\mathbf{10}$ allows 
spontaneous GUT symmetry breaking to be achieved using a $\mathbf{10}$ and 
a $\mathbf{\bar{10}}$ of superheavy Higgs fields with the same and 
opposite U(1) hypercharges. Their neutral components 
develop large vacuum expectation values ({\it vev}'s) breaking SU(5) 
$\times$ U(1) $\to$ SU(3) $\times$ SU(2) $\times$ U(1), while the 
electroweak spontaneous breaking occurs through the Higgs doublets 
$\mathbf{H}_2$ and $\mathbf{H}_{\bar{2}}$:
\begin{equation}
h_{\mathbf{5}} = \left\{\mathbf{H}_2,\mathbf{H}_3\right\}\ \ \  
; \ \ \ h_{\mathbf{\bar{5}}} = \left\{\mathbf{H_{\bar{2}}}, 
\mathbf{H_{\bar{3}}} \right\}.
\end{equation}
The resulting economical doublet-triplet splitting mechanism 
gives a large mass to the Higgs triplets 
$(\mathbf{H}_3,\mathbf{H_{\bar{3}}})$ by coupling them to states
in $\mathbf{10}$ and $\mathbf{\bar{10}}$ Higgs representations,
through trilinear superpotential couplings of the form 
\begin{eqnarray}
 FFh \rightarrow d^c_H\left\langle \nu^c_H\right\rangle H_3 \\
 \bar{F}\bar{F}\bar{h} \rightarrow \bar{d}^c_H\left\langle 
\bar{\nu}^c_H\right\rangle H_{\bar{3}} ,
\end{eqnarray}
while keeping light the Higgs doublets $(\mathbf{H}_2, 
\mathbf{H_{\bar{2}}})$ light.

The absence of any mixing between the Higgs triplets 
$(\mathbf{H}_3,\mathbf{H_{\bar{3}}})$ in
this dynamic doublet-triplet splitting mechanism provides a natural 
suppression of the $d=5$ operators that might
otherwise mediate rapid proton decay, so that flipped SU(5) is probably 
the simplest unified model that can satisfy the experimental limits 
placed on the proton lifetime~\cite{Ellis:2002vk}.  

String-derived flipped SU(5) belongs to a class of models constructed
using free fermions on the world sheet, corresponding to compactification
on the $Z_2\times Z_2$ orbifold at the maximally-symmetric point in the
Narain moduli space~\cite{Narain}.
Although this model was originally constructed in the weak-coupling limit,
it is possible that it may be elevated in the strong-coupling limit to an
authentic $M$-theory model and may at some point make contact with models
based on D-brane constructions.

The full gauge group of the model is SU(5) $\times$ U(1) $\times$ U(1)$^4 
\times$ SO(10) $\times$ SO(6) [$\simeq$ SU(4)], where the latter two factors 
are confining hidden-sector gauge interactions. The matter spectrum 
comprises the following fields: \\
(i) {\it Observable sector:} 
The conventional matter fields may be regarded as three $\mathbf{16}$ 
representations of SO(10) that $\supset$
SU(5) $\times$ U(1) chiral multiplets $F_i 
(\mathbf{10},\frac{1}{2}),\
\overline{f}_i(\overline{\mathbf 5}, -\frac{3}{2}),\
l^c_i(\mathbf{1},\-\frac{5}{2}) (i = 1, 2, 3)$; extra matter fields
$F_4(\mathbf{10},\frac{1}{2})$, $f_4(\mathbf{5},\frac{3}{2})$,
$\bar{l}^c_4(\mathbf{1},-\frac{5}{2})$ and
$\bar{F}_5(\overline{\mathbf{10}}, -\frac{1}{2})$,
$\bar{f}_5(\bar{\mathbf{5}}, -\frac{3}{2})$, $l^c_5(\mathbf{1},
\frac{5}{2})$; and four Higgs-like fields in the $\mathbf{10}$
representation of SO(10), that $\supset$ 
$h_i(\mathbf{5}, -1)$, $\bar{h}_i(\bar{\mathbf{5}},1)$, $i = 1, 2, 3, 
45$.
In our realization of the model, we make the following
flavour identifications of the Standard Model states 
with the various string representations:
\begin{eqnarray}
t \ b \ \tau \ \nu_{\tau}: \ \ \  Q_4 \ d^c_4 \ u^c_5 \ L_1 \ l^c_1, \\
\nonumber c \ s \ \mu \ \nu_{\mu}: \ \ \ Q_2 \ d^c_2 \ u^c_2 \ L_2 \ 
l^c_2, \\
\nonumber u \ d \ e \ \nu_{e}: \ \ \ Q_{\beta} \ d^c_{\beta} \ u^c_1 \ 
L_5 \ l^c_5 ,  
\end{eqnarray}
where $\beta$ indicates a mixture of fields with the indices $1$ and 
$3$. The light Higgs doublets $h_d$ and $h_u$ (where 
$h_d$ is the Higgs doublet responsible for giving mass to the down-type 
quarks and leptons and $h_u$ gives mass to the up-type quarks)
are contained in in the 
$h_1$ and $\bar{h}_{45}$ string pentaplet representations, respectively. 
\\
(ii) {\it Singlets:}  
There are ten gauge-singlet fields $\phi_{45}$, $\phi^+$, $\phi^-$,
$\phi_i (i = 1, 2, 3, 4)$, $\Phi_{12}$, $\Phi_{23}$, $\Phi_{31}$, their
ten \lq barred\rq \ counterparts, and five extra fields $\Phi_I (I = 1
\cdots 5)$. \\
(iii) {\it Hidden sector:}  
This contains 22 matter fields in the following representations of 
$SO(10)_h \otimes SU(4)_h$:
$T_i(\mathbf{10},\mathbf{1})$, $\Delta_i(\mathbf{1}, \mathbf{6}) (i = 1
\cdots 5)$; $\tilde{F}_i(\mathbf{1}, \mathbf{4})$,
$\tilde{\bar{F}}_i(\mathbf{1}, \mathbf{\bar{4}})(i = 1 \cdots 6)$. Flat
potential directions along which the anomalous combination of 
hypercharges U(1)$_A$ is cancelled
induce masses that are generally near the string scale
for some, but not all, of these states.  Depending upon the
number of $T_i$ and $\Delta_i$ states remaining massless, the SO(10)
condensate scale is $10^{14-15}$ GeV and the SU(4) condensate scale is 
$10^{11-13}$~GeV~\cite{Lopez:1995cs}. 
The $\tilde{F}_{3,5}$ and $\tilde{\bar{F}}_{3,5}$ states always
remain massless down to the condensate scale.  The U(1)$_i$ charges and
hypercharge assignments are shown in the Table below.
 
\vskip 0.3cm
{\centering \begin{tabular}{|c|c|} \hline
$\Delta^0_1(0,1,6,0,-\frac{1}{2},\frac{1}{2},0)$&
$\Delta^0_2(0,1,6,-\frac{1}{2},0,\frac{1}{2},0)$\\
$\Delta^0_3(0,1,6,-\frac{1}{2},-\frac{1}{2},0,\frac{1}{2})$&
$\Delta^0_4(0,1,6,0,-\frac{1}{2},\frac{1}{2},0)$\\
$\Delta^0_5(0,1,6,\frac{1}{2},0,-\frac{1}{2},0)$& \\
$T^0_1(10,1,0,-\frac{1}{2},\frac{1}{2},0)$ &
$T^0_2(10,1,-\frac{1}{2},0,\frac{1}{2},0)$ \\
$T^0_3(10,1,-\frac{1}{2},-\frac{1}{2},0,\frac{1}{2})$ &
$T^0_4(10,1,0,\frac{1}{2},-\frac{1}{2},0)$ \\
$T^0_5(10,1,-\frac{1}{2},0,\frac{1}{2},0)$&   \\ \hline 
\end{tabular}\par}
\vspace*{0.3 cm}
 
{\centering \begin{tabular}{|c|c|} \hline ${\tilde
F}^{+\frac{1}{2}}_1(1,4,-\frac{1}{4},\frac{1}{4},-\frac{1}{4},\frac{1}{2})$
& ${\tilde
F^{+\frac{1}{2}}}_2(1,4,-\frac{1}{4},\frac{1}{4},-\frac{1}{4},-\frac{1}{2})$
\\ ${\tilde
F}^{-\frac{1}{2}}_3(1,4,\frac{1}{4},\frac{1}{4},-\frac{1}{4},\frac{1}{2})$
& ${\tilde
F}^{+\frac{1}{2}}_4(1,4,\frac{1}{4},-\frac{1}{4},-\frac{1}{4},6-\frac{1}{2})$
\\ ${\tilde
F}^{+\frac{1}{2}}_5(1,4,-\frac{1}{4},\frac{3}{4},-\frac{1}{4},0)$ &
${\tilde
F}^{+\frac{1}{2}}_6(1,4,-\frac{1}{4},\frac{1}{4},-\frac{1}{4},
-\frac{1}{2})$\\ ${\tilde {\bar
F}}^{-\frac{1}{2}}_1(1,4,-\frac{1}{4},\frac{1}{4},\frac{1}{4},\frac{1}{2})$
& ${\tilde {\bar
F}}^{-\frac{1}{2}}_2(1,4,-\frac{1}{4},\frac{1}{4},\frac{1}{4},-\frac{1}{2})$
\\ ${\tilde {\bar
F}}^{+\frac{1}{2}}_3(1,4,-\frac{1}{4},-\frac{1}{4},\frac{1}{4},-\frac{1}{2})$
& ${\tilde {\bar
F}}^{-\frac{1}{2}}_4(1,4,-\frac{1}{4},\frac{1}{4},\frac{1}{4},
-\frac{1}{2})$ \\ ${\tilde {\bar
F}}^{-\frac{1}{2}}_5(1,4,-\frac{3}{4},\frac{1}{4},-\frac{1}{4},0)$ &
${\tilde {\bar
F}}^{-\frac{1}{2}}_6(1,4,\frac{1}{4},-\frac{1}{4},\frac{1}{4},
-\frac{1}{2})$\\ \hline
\end{tabular}\par}
\vspace*{0.3 cm}
Table:  {\it The spectrum of hidden matter fields 
that are massless at the string scale in the revamped flipped SU(5) model.
We display the quantum numbers under the hidden 
gauge group SO(10) $\times$
SO(6) $\times$ U(1)$^4$, and subscripts indicate the electric charges.}
{}\\

The hidden-sctor matter fields are confined into crypton bound states. 
These occur in `cryptospin' multiplets with different
permutations of confined constituents, analogous to the flavour SU(2), 
SU(3) and SU(4) multiplets of bound states in QCD. The cryptospin 
multiplets contain doubly-charged tetrons
\begin{equation}
\Psi^{--} = \tilde{F}_3\tilde{F}_3\tilde{F}_3\tilde{F}_3, \; \;
\Psi^{++} = \tilde{F}_5\tilde{F}_5\tilde{F}_5\tilde{F}_5,
\end{equation}
\begin{equation}
\bar{\Psi}^{++} = 
\tilde{\bar{F}}_3\tilde{\bar{F}}_3\tilde{\bar{F}}_3\tilde{\bar{F}}_3, 
\; 
\;  \bar{\Psi}^{--} = 
\tilde{\bar{F}}_5\tilde{\bar{F}}_5\tilde{\bar{F}}_5\tilde{\bar{F}}_5,
\end{equation}
and singly-charged tetrons
\begin{equation}
\Psi^{+} = \tilde{F}_3\tilde{F}_5\tilde{F}_5\tilde{F}_5, \; \;
\Psi^{-} = \tilde{F}_5\tilde{F}_3\tilde{F}_3\tilde{F}_3,
\end{equation}
\begin{equation}
\bar{\Psi}^{-} = 
\tilde{\bar{F}}_3\tilde{\bar{F}}_5\tilde{\bar{F}}_5\tilde{\bar{F}}_5, 
\; 
\;
\bar{\Psi}^{+} = 
\tilde{\bar{F}}_5\tilde{\bar{F}}_3\tilde{\bar{F}}_3\tilde{\bar{F}}_3,
\end{equation}
as well as neutral tetrons 
\begin{equation}
\Psi^{0} = \tilde{F}_3\tilde{F}_3\tilde{F}_5\tilde{F}_5, \; \;
\bar{\Psi}^{0} = 
\tilde{\bar{F}}_3\tilde{\bar{F}}_3\tilde{\bar{F}}_5\tilde{\bar{F}}_5.
\end{equation}
We have shown previously that the charged tetrons could have decayed with 
rather short lifetimes early in the history of the 
Universe~\cite{EMN1}, and so avoid any problems with charged dark 
matter particles~\cite{Alon}. 

However, the neutral tetrons can decay only via higher-order operators in
the superpotential, which we have identified in~\cite{EMN1}, and
may have lifetimes exceeding the age of the Universe. This may make them
good candidates for cold dark matter, and their slow decays are a possible
source of the UHECRs.

In the next Section, we turn our attention to these interactions and
examine the UHECR energy spectra that may result from these specific
tetron decays.

\section{UHECR Injection Spectra}

We have previously found ~\cite{EMN1} the following 10$^{th}$-order 
superpotential operators through which the neutral tetrons may decay:
\begin{eqnarray}
{\Psi}^0   [F_2 F_2 \bar{\Phi}_{31} \bar{\phi}_{45} \phi^- h_1 + 
F_2 F_2 \Phi_{23} \bar{\phi}_{45} \bar{\phi}^+  h_1 +
F_2 F_3 F_3 \phi_4  \bar{\phi}_{45} \bar{f}_2 + \\
\nonumber
F_4 \Phi_{23} \bar{\phi}_{45} \phi^- \bar{h}_{45} \bar{f}_5 +   
(\bar{\Phi}_{31} \bar{\phi}_{45} \phi^-   \\
\nonumber
+ \Phi_{23} \bar{\phi}_{45} \bar{\phi}^+)h_1(\bar{f}_2 l^c_2 + 
\bar{f}_5 l^c_5) 
+ \nonumber \Phi_{23} \bar{\phi}_{45} \phi^- h_1 \bar{f}_1 l^c_1]. 
\end{eqnarray}
\begin{eqnarray}
\bar{\Psi}^0  [F_2 F_2 \Phi_{31} \phi_{45} \bar{\phi}^- h_1 + 
F_2 F_2 \bar{\Phi}_{23} \phi_{45} \phi^+ h_1 + 
F_2 F_2 \bar{\phi}^- h_1 h_1 \bar{h}_{45} + \\
\nonumber
F_4 F_4 \Phi_{31} \phi_{45} \phi^+ h_1 +  F_4 F_4 \phi^+ h_1 h_1 \bar{h}_{45} + 
F_4 \Phi_{31} \phi_{45} \phi^+ \bar{h}_{45} \bar{f}_5 + \\
\nonumber
F_4 \phi^+ h_1 \bar{h}_{45} \bar{h}_{45} \bar{f}_{5} + F_4 \bar{\phi}^- 
h_1 h_1 h_1 l^c_5 + \\
\nonumber (\Phi_{31} \phi_{45} \phi^+ h_1 + \phi^+ h_1 h_1 
\bar{h}_{45})\bar{f}_1 l^c_1 + 
(\Phi_{31} \phi_{45} \bar{\phi}^- h_1 + \bar{\Phi}_{23} \phi_{45} 
\phi^+ h_1 + \\
\nonumber
\phi^+ \bar{h}_{45} \bar{h}_{45} \bar{h}_{45} + 
\bar{\phi}^- h_1 h_1 \bar{h}_{45})(\bar{f}_2 l^c_2 + \bar{f}_5 l^c_5)] .
\end{eqnarray}
Using the flavour identifications we outlined above, these operators would 
give rise to the following neutral tetron decay modes:
\begin{eqnarray}
\Psi^0 \rightarrow \tau \ \tau^c \ h_d \ \phi^3, 
\Psi^0 \rightarrow e/\mu \ e^c/\mu^c \ h_d \ \phi^3, 
\Psi^0 \rightarrow b \ b^c \ h_d \ \phi^3, \\
\nonumber
\Psi^0 \rightarrow b \ b^c \ h_d \ h_d \ h_u \phi, 
\Psi^0 \rightarrow t \ t^c \ h_u \ \phi^3, 
\Psi^0 \rightarrow t \ t^c \ h_u \ h_u \ h_d \ \phi, \\
\nonumber
\Psi^0 \rightarrow c \ c^c \ d \ d^c \ \phi^2, 
\Psi^0 \rightarrow s \ s^c \ h_d \ \phi^3.
\label{possdex}
\end{eqnarray}
We note that there are several different possible decay modes,
any of which may be dominant, depending on unknown features of
the model dynamics that determine the relative values of their 
coefficients. In particular, the most important tetron
decays could be into either leptons or quarks, and there are many
different possibilities for the dominant flavours.

We plot in Figs.~1 to 8 below the expected UHECR energy spectra of 
photons and nucleons due to each of these possible tetron decay modes, as 
well as the maximum photon fractions expected. 
The energy spectra were calculated for a mass $M_X = 
2\cdot10^{13}$~GeV, using 
the fragmentation functions 
$D^i(x,M_X^2)$ generated by the code \texttt{SHdecay}~\cite{Barbot:2003wv}. 
This code 
calculates the fragmentation into the seven stable MSSM particles 
($p$, $\gamma$, $e$, neutralino LSP $\chi$, $\nu_e$, $\nu_{\mu}$, 
$\nu_{\tau}$) for any given initial decay parton.  
The many-body decays distribute the total decay energy $M_X$ among the 
different particles. We include Higgs decays, but we ignore the decays of 
the singlet fields, except to take into account
their kinematical effects on the primary quark and lepton spectra.
We follow~\cite{Sarkar:2001se} in
estimating the probability density $\rho_n(z)$ that one decay 
parton carries off a fraction $z$ of the total available decay energy
$M_X$:
\begin{equation}
\rho_n(z)=(n-1)(n-2)z(1-z)^{n-3}
\end{equation}
for $n \geq 3$ decay partons. The resulting flux from the emission of a given decay parton is then 
\begin{equation}
E^3J^i(E)=  
Bx^3\int^{1}_{x}\frac{dz}{z}\rho_n\left(\frac{x}{z}\right)D^i(z,M_x^2),
\label{convolve}
\end{equation}
where $i$ = ($p$, $\gamma$, $e$, $\chi$, $\nu_e$, $\nu_{\mu}$, 
$\nu_{\tau}$).

To obtain the total UHECR spectrum, we add to this the background flux of
nucleons that would be expected to result from a homogenous distribution of
extragalactic sources that exhibits the distinctive pile-up due to the GZK
effect~\cite{Berezinsky:2005cq}~\footnote{However, we note that this model
is likely to come under pressure from upper limits on high-energy
cosmic-ray neutrinos~\cite{Ahlers} - private communication from S.~Sarkar,
see also {\tt
http://www-thphys.physics.ox.ac.uk/users/SubirSarkar/talks/munich05.pdf}.}.  
The constant $B$ in (\ref{convolve}) is a normalization coefficent
determined by the tetron number density and lifetime, \textit{viz}
\begin{equation}
B \sim R^{halo} \frac{n_X}{\tau_X} \frac{1}{M_X}
\end{equation}
This dimensional coefficient $B$ is
not determined {\it a priori}, and must be fitted to the experimental data.  
In each of Figs.~1 to 8, we show the total spectrum
obtained by summing the background and the fluxes of nucleons and
photons resulting from tetron decay, and in a second panel we display the
gamma fractions: $\gamma / ( \gamma + p)$. We have assumed no photon
attentuation in the calculated spectra, although a strong attenuation
cannot be excluded~\cite{SarkarAPP}, because the galactic radio background
has never been accurately measured and its intensity is largely
unknown~\cite{Protheroe:1996si}.

\begin{figure}[f]
  \centering
	 \begin{minipage}[c]{0.45\textwidth}
	  \fbox{\includegraphics[width=0.9\textwidth]{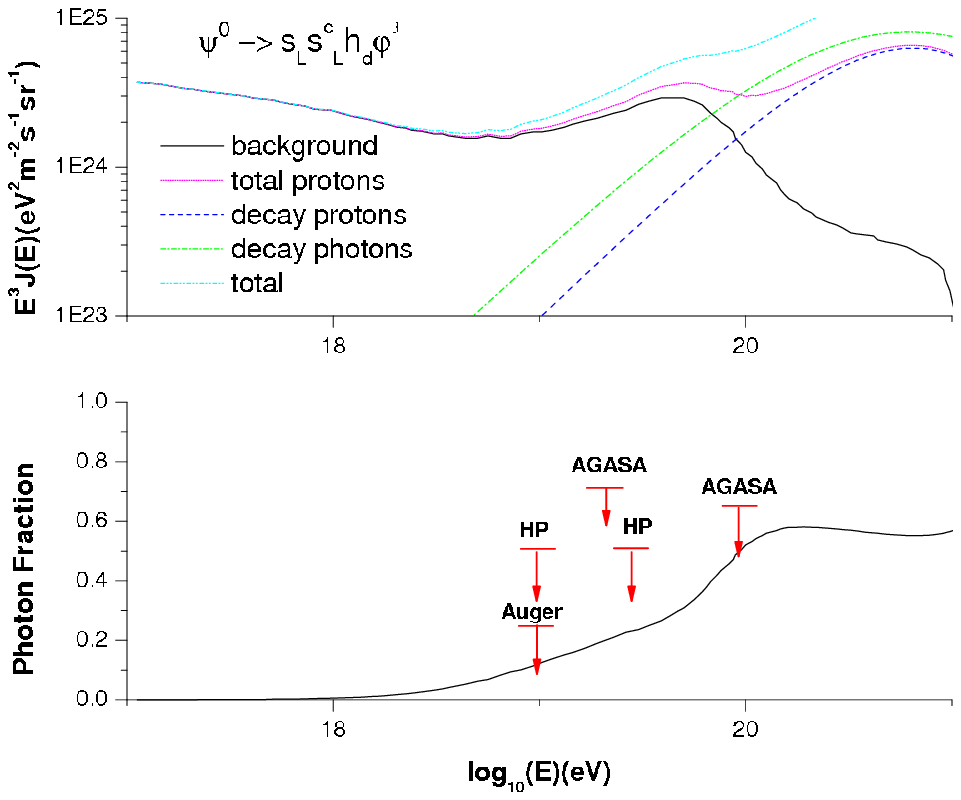}}
	  \caption{
The top panel shows the total UHECR spectrum and the bottom panel the 
photon fraction from the decay mode 
$\Psi^0 \rightarrow s \ s^c \ h_d \ \phi^3$.}
	 \end{minipage}
\hspace{0.5cm}
	 \begin{minipage}[c]{0.45\textwidth}
	  \fbox{\includegraphics[width=0.9\textwidth]{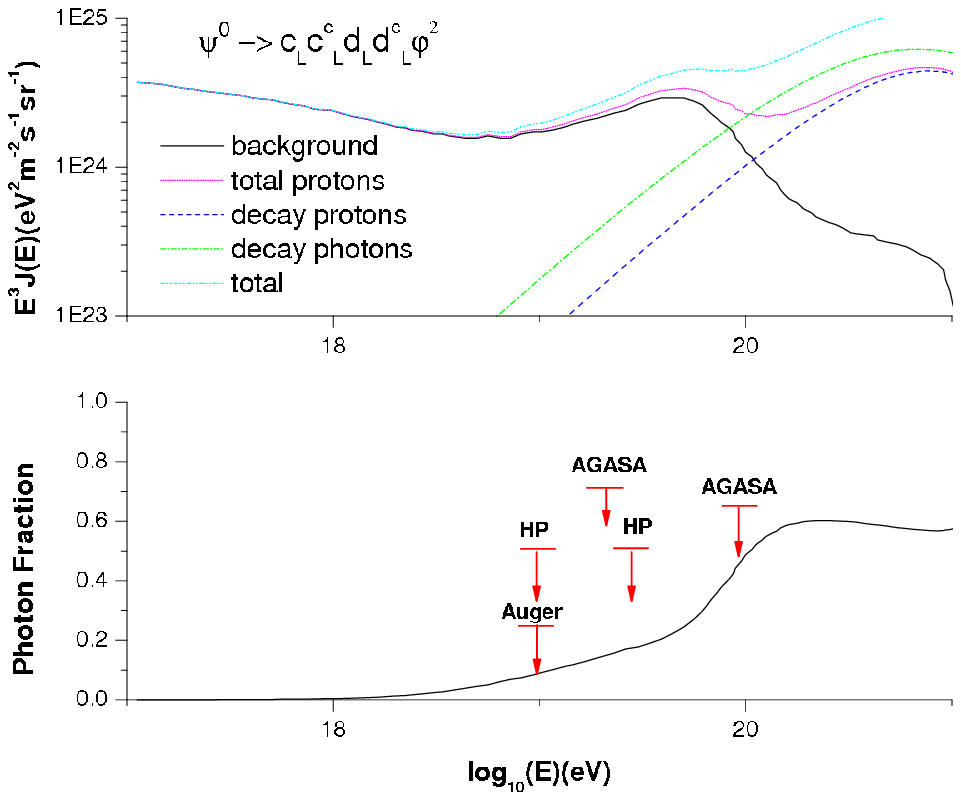}}
	  \caption{
The top panel shows the total UHECR spectrum and the bottom panel the 
photon fraction from the decay mode 
$\Psi^0 \rightarrow c \ c^c \ d \ d^c \ \phi^2$.}
	 \end{minipage}%
\end{figure}

\begin{figure}[f]
   \centering
	 	 \begin{minipage}[c]{0.45\textwidth}
\fbox{\includegraphics[width=0.9\textwidth]{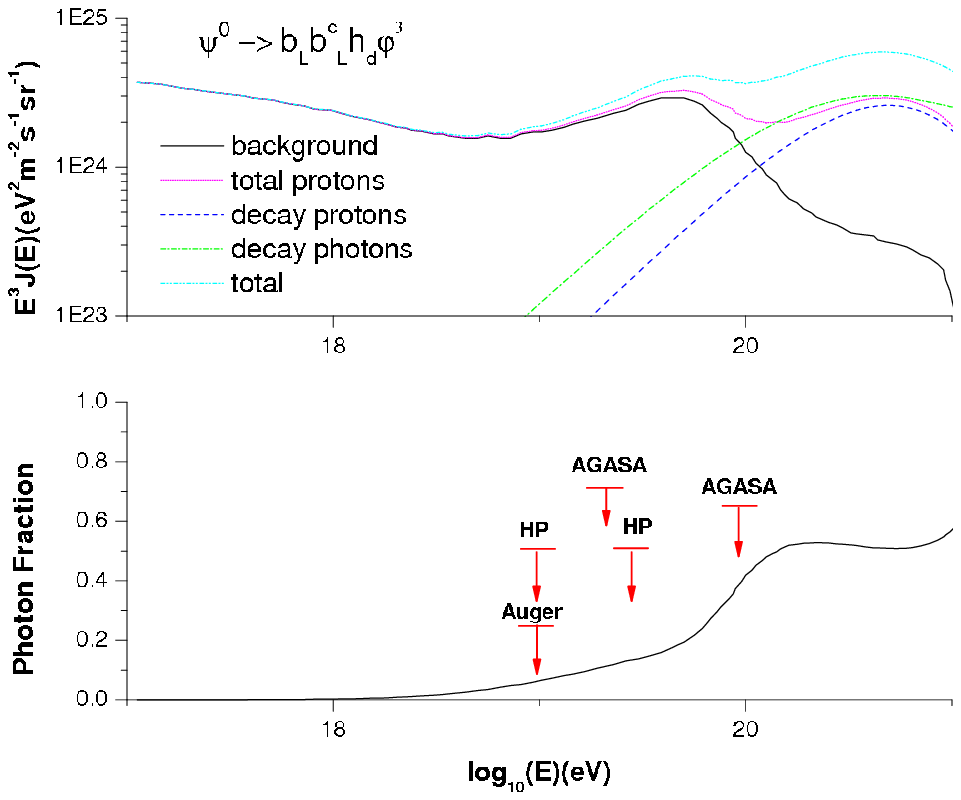}}
	  \caption{
The top panel shows the total UHECR spectrum and the bottom panel the 
photon fraction from the decay mode 
$\Psi^0 \rightarrow b \ b^c \ h_d \ \phi^3$.}
	 \end{minipage}%
\hspace{0.5cm}
	 \begin{minipage}[c]{0.45\textwidth}
	  \fbox{\includegraphics[width=0.9\textwidth]{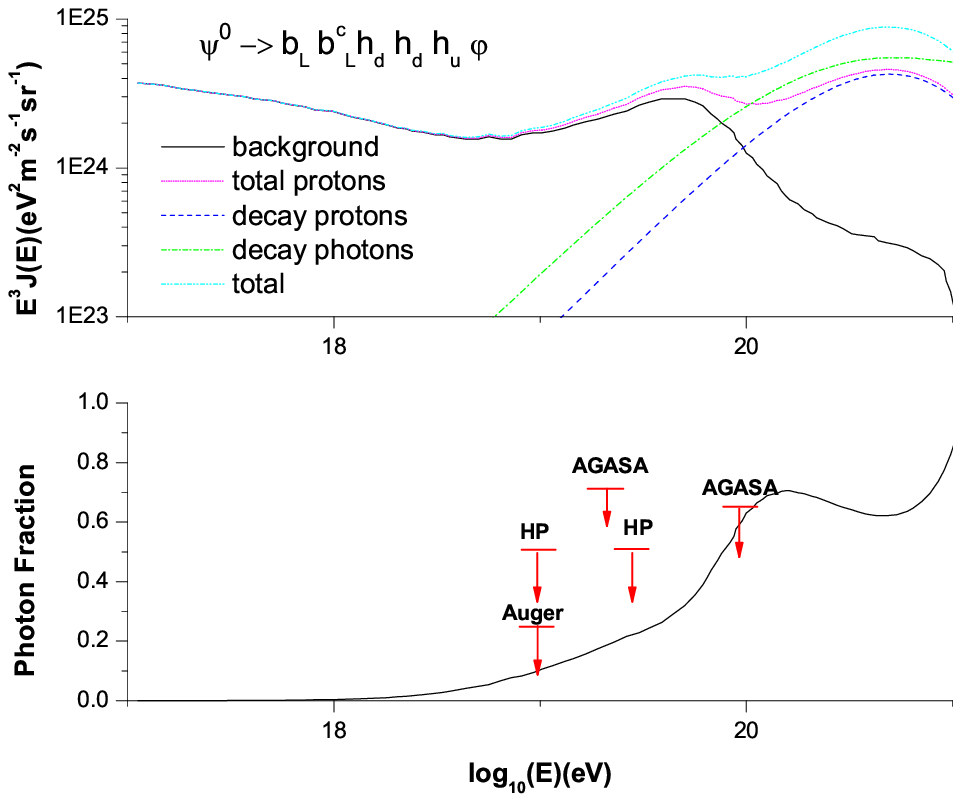}}
	  \caption{
The top panel shows the total UHECR spectrum and the bottom panel the 
photon fraction from the decay mode 
$\Psi^0 \rightarrow b \ b^c \ h_d \ h_d \ h_u \phi$.}
	 \end{minipage}
\end{figure}

\begin{figure}[f]
  \centering
	 \begin{minipage}[c]{0.45\textwidth}
	  \fbox{\includegraphics[width=0.9\textwidth]{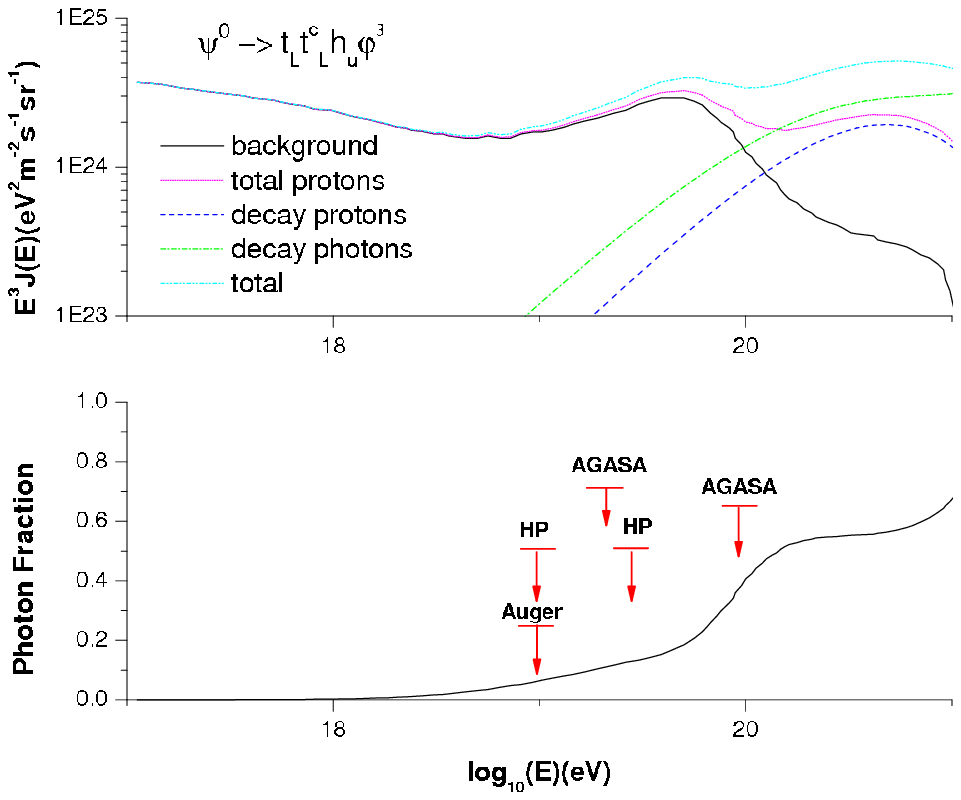}}
	  \caption{
The top panel shows the total UHECR spectrum and the bottom panel the 
photon fraction from the decay mode 
$\Psi^0 \rightarrow t \ t^c \ h_u \ \phi^3$.}
	 \end{minipage}%
\hspace{0.5cm}
	 \begin{minipage}[c]{0.45\textwidth}
	  \fbox{\includegraphics[width=0.9\textwidth]{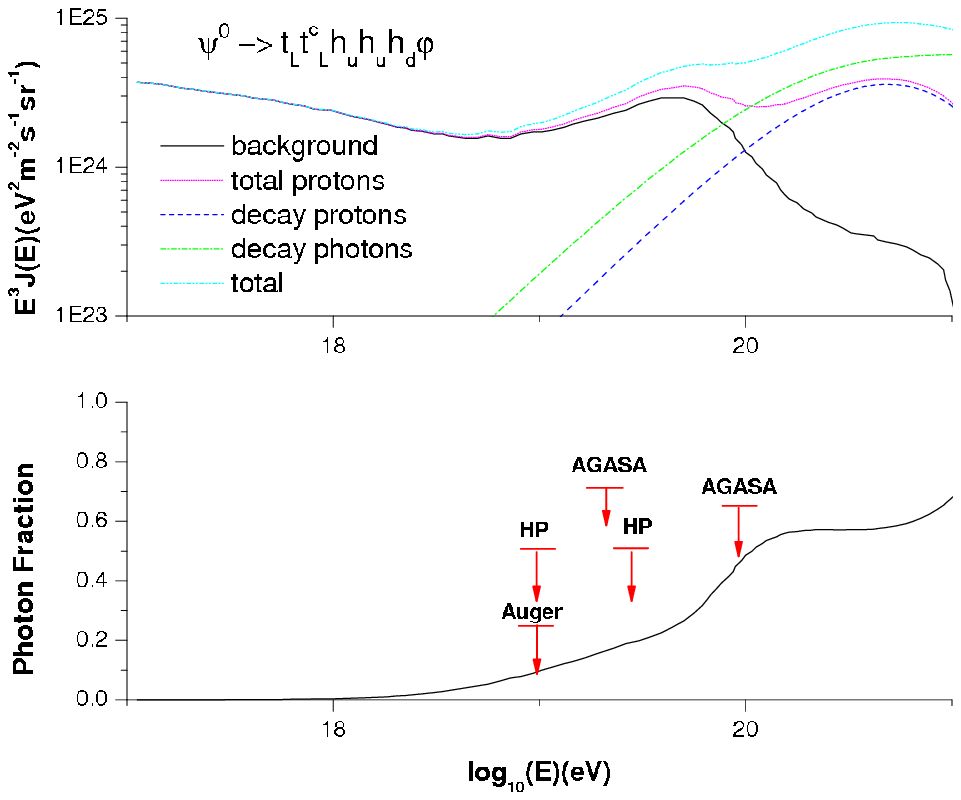}}
	  \caption{
The top panel shows the total UHECR spectrum and the bottom panel the 
photon fraction from the decay mode 
$\Psi^0 \rightarrow t \ t^c \ h_u \ h_u \ h_d \ \phi$.}
	 \end{minipage}
\end{figure}

\begin{figure}[f]
	\centering
	 \begin{minipage}[c]{0.45\textwidth}
	  \fbox{\includegraphics[width=0.9\textwidth]{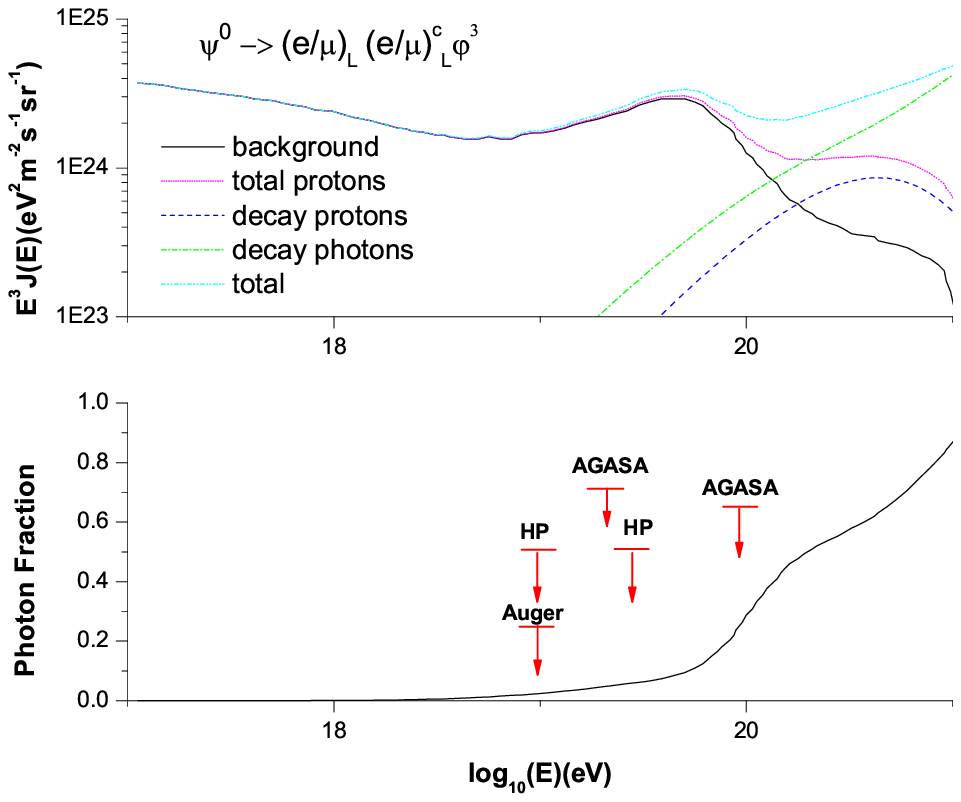}}
	  \caption{
The top panel shows the total UHECR spectrum and the bottom panel the 
photon fraction from the decay mode 
$\Psi^0 \rightarrow e/\mu \ e^c/\mu^c \ h_d \ \phi^3$.}
	 \end{minipage}
\hspace{0.5cm}
	 \begin{minipage}[c]{0.45\textwidth}
		\fbox{\includegraphics[width=0.9\textwidth]{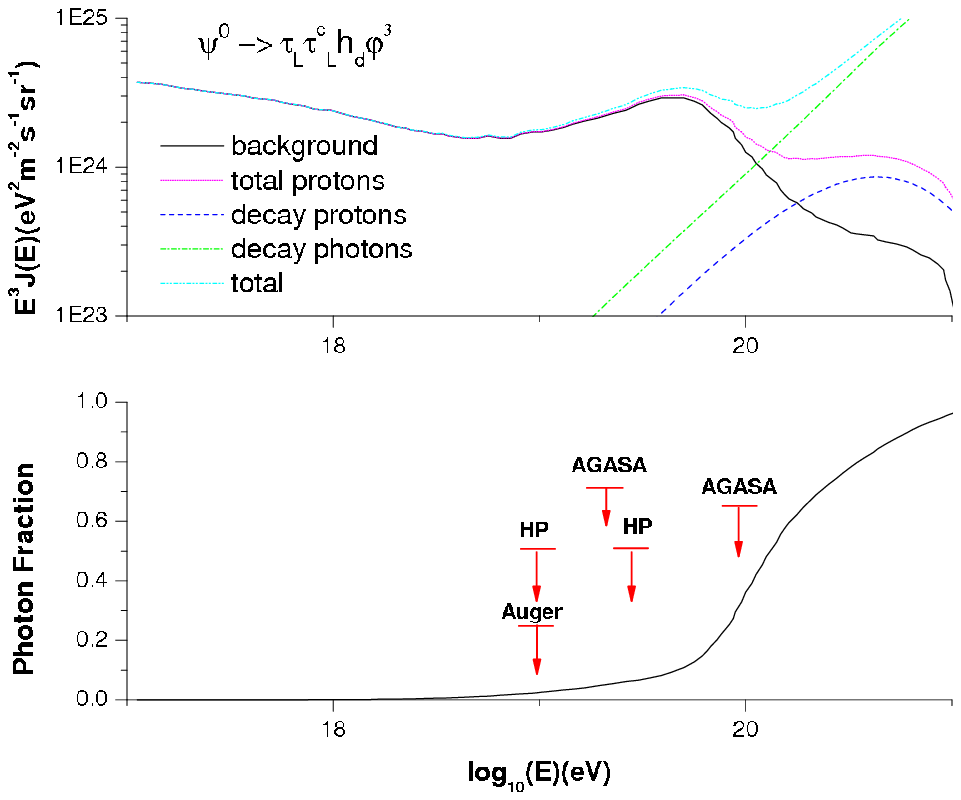}}
		\caption{
The top panel shows the total UHECR spectrum and the bottom panel the 
photon fraction from the decay mode 
$\Psi^0 \rightarrow \tau \ \tau^c \ h_d \ \phi^3$.}
	 \end{minipage}%
  \end{figure}	 

Figs.~1 to 6 are for quark primaries, and are ordered according to the
masses of the quarks involved. In the case of the $b$ quark, we show in
Figs.~3 and 4 plots for two superpotential operators with different
numbers of accompanying Higgs fields: the two plots are rather similar,
and the same is true of the two plots shown in Figs.~5 and 6 for primary
$t$ quarks. We are thus led to hope that including the (model-dependent)
decays of the singlets $\phi$ would not have large effects.  The plots for
lepton primaries shown in Figs.~7 and 8 are more distinctive, in that the
photon fractions rise to much larger values at energies above
$10^{20}$~eV~\footnote{The photon fractions for second-generation quark
primaries look somewhat flatter than those for $b$ and $t$ quarks, but
this difference is probably within the modelling uncertainties.}.

In Fig.~9 we compare the spectrum for one of the operators with primary
$b$ quarks, calculated for a crypton mass of $10^{13}$~GeV, 
with experimental data from the Fly's Eye,
HiRes, AGASA, and Auger experiments~\cite{Nagano:2000ve,HiRes,AGASA,Sommers:2005vs}. 
The AGASA flux has been scaled by a
factor of 0.55 for consistency with the other data, and the normalizations
for the crypton decay contributions to these spectra has been adjusted for
the different crypton masses. The limited statistics for UHECRs with
energies $\geq 10^{19}$~eV available in the present data sets do not offer
any clear discrimination between crypton masses in the range $ 2 \times
10^{13}~{\rm GeV} \geq M_X \geq 10^{12}$~GeV.  In the case 
of a crypton mass $\sim 10^{12}$~GeV there is no clear signal of 
a crypton contribution to the UHECR since the flux from such a decay
is essentially buried within the background from homogenous extragalatic sources.
A clear signal of crypton decay, at least in this model, would require
a lower limit on the crypton mass $M_X \geq 5\cdot10^{12}$~GeV in order
to provide an excess of events above $4\cdot10^{19}$~eV that could not be
attributable to extragalactic astrophysical acceleration mechanisms.  

\begin{figure}[t]
  \centering
   \begin{minipage}[c]{0.65\textwidth}
\fbox{\includegraphics[width=0.9\textwidth]{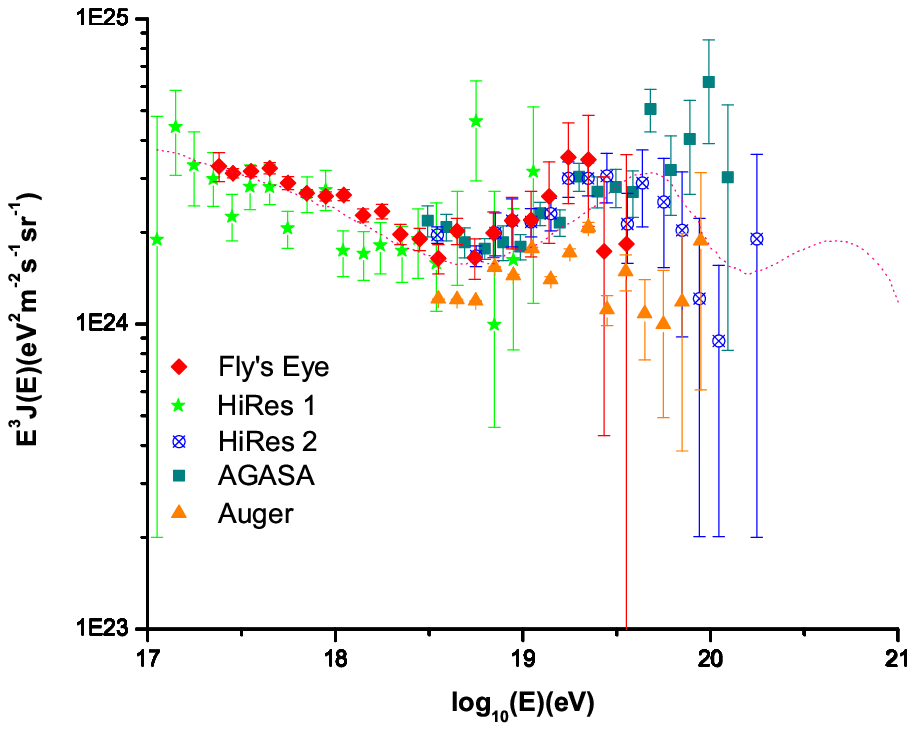}}
    \caption{
A comparison with the available data on the UHECRs from the Fly's Eye, 
HiRes, AGAS and Auger experiments with the crypton decay model
$\Psi^0 \rightarrow b \ b^c \ h_d \ \phi^3$ for $M_X = 10^{13}$~GeV.}
   \end{minipage}
\end{figure}  

\section{Conclusions}

We have carried as far as is possible at present the modelling of flipped
crypton decay contributions to UHECRs, including all the possible
$10^{th}$-order superpotential operators. The experimental data presently
available are consistent with all the decay modes possible in this crypton
framework. The total UHECR spectra are consistent with a contribution from
cryptons weighing between $ 2 \times 10^{13}~{\rm GeV}$ and $10^{12}$~GeV, 
although only a crypton mass $M_X \geq 5\cdot 10^{12}$~GeV would provide
an unambiguous signal over conventional explanations.
The available upper limits on the possible photon fraction do not
exclude any of the crypton models we have studied.

In the future, the larger data set expected from Auger may be able to 
discriminate between crypton decays and other models of UHECRs, and also 
among different crypton models themselves. Greater statistics will enable 
the UHECR anisotropy to be measured with sufficient accuracy to 
discriminate crypton decay from a uniform distribution of astrophysical 
sources, and more accurate measurements of the photon fraction at higher 
energies might offer some discrimination between models with lepton and 
quark primaries, as seen by comparing Figs.~1 to 6 with Figs.~7 and 8 
above.

Thus there is hope that, in the near future, we may finally learn whether
UHECRs have a macrophysical origin or a microphysical origin and, in the
latter case, may start to discriminate between different microphysical
models.

\section*{Acknowledgements}

The work of D.V.N. is supported by DOE grant DE-FG03-95-ER-40917.


\begin{thebibliography}{99}

\bibitem{Greisen:1966jv}
  K.~Greisen,
  Phys.\ Rev.\ Lett.\  {\bf 16}, 748 (1966).
  
\bibitem{Zatsepin:1966jv}
  G.~T.~Zatsepin and V.~A.~Kuzmin,
  JETP Lett.\  {\bf 4}, 78 (1966)
  [Pisma Zh.\ Eksp.\ Teor.\ Fiz.\  {\bf 4}, 114 (1966)].

\bibitem{Nagano:2000ve}
  M.~Nagano and A.~A.~Watson,
  Rev.\ Mod.\ Phys.\  {\bf 72}, 689 (2000).

\bibitem{Westerhoff}
For a recent review, see:
S.~Westerhoff, hep-ex/0512018.

\bibitem{AGASA}
  M.~Takeda {\it et al.},
  Phys.\ Rev.\ Lett.\  {\bf 81}, 1163 (1998)
  [arXiv:astro-ph/9807193].
  M.~Takeda {\it et al.},
  Astropart.\ Phys.\  {\bf 19}, 447 (2003)
  [arXiv:astro-ph/0209422].
  
\bibitem{Evans}
  N.~W.~Evans, F.~Ferrer and S.~Sarkar,
  Astropart.\ Phys.\  {\bf 17}, 319 (2002)
  [arXiv:astro-ph/0103085].

\bibitem{HiRes}
  R.~U.~Abbasi {\it et al.}  [High Resolution Fly's Eye Collaboration],
  Phys.\ Rev.\ Lett.\  {\bf 92}, 151101 (2004)
  [arXiv:astro-ph/0208243];
  R.~U.~Abbasi {\it et al.}  [The High Resolution Fly's Eye Collaboration],
  Phys.\ Lett.\ B {\bf 619}, 271 (2005)
  [arXiv:astro-ph/0501317].
 
\bibitem{Sommers:2005vs}
  P.~Sommers  [Pierre Auger Collaboration],
  arXiv:astro-ph/0507150.

\bibitem{inflation}
  E.~W.~Kolb, A.~D.~Linde and A.~Riotto,
  Phys.\ Rev.\ Lett.\  {\bf 77}, 4290 (1996)
  [arXiv:hep-ph/9606260];
  B.~R.~Greene, T.~Prokopec and T.~G.~Roos,
  Phys.\ Rev.\ D {\bf 56}, 6484 (1997)
  [arXiv:hep-ph/9705357];
  E.~W.~Kolb, A.~Riotto and I.~I.~Tkachev,
  Phys.\ Lett.\ B {\bf 423}, 348 (1998)
  [arXiv:hep-ph/9801306];
  D.~J.~H.~Chung, E.~W.~Kolb and A.~Riotto,
  Phys.\ Rev.\ D {\bf 59}, 023501 (1999)
  [arXiv:hep-ph/9802238].
  
\bibitem{Schellekens:1989qb}
  A.~N.~Schellekens,
  Phys.\ Lett.\ B {\bf 237}, 363 (1990).
  
\bibitem{cryptons}
  J.~R.~Ellis, J.~L.~Lopez and D.~V.~Nanopoulos,
  Phys.\ Lett.\ B {\bf 247}, 257 (1990);
  K.~Benakli, J.~R.~Ellis and D.~V.~Nanopoulos,
  Phys.\ Rev.\ D {\bf 59}, 047301 (1999)
  [arXiv:hep-ph/9803333].

\bibitem{EMN1}
  J.~R.~Ellis, V.~E.~Mayes and D.~V.~Nanopoulos,
  Phys.\ Rev.\ D {\bf 70}, 075015 (2004)
  [arXiv:hep-ph/0403144].
  
\bibitem{Risse:2005hi}
  M.~Risse  [Pierre Auger Collaboration],
  arXiv:astro-ph/0507402.
  
\bibitem{Sarkar}
S.~Sarkar, private communication.
  
\bibitem{Risse:2005jr}
  M.~Risse {\it et al.},
  Phys.\ Rev.\ Lett.\  {\bf 95}, 171102 (2005)
  [arXiv:astro-ph/0502418].
  
\bibitem{Barbot:2003wv}
  C.~Barbot,
  arXiv:hep-ph/0308028.

\bibitem{Berezinsky:1997hy}
  V.~Berezinsky, M.~Kachelriess and A.~Vilenkin,
  Phys.\ Rev.\ Lett.\  {\bf 79}, 4302 (1997)
  [arXiv:astro-ph/9708217].

\bibitem{Kuzmin:1997cm}
  V.~A.~Kuzmin and V.~A.~Rubakov,
   ``Ultrahigh-energy cosmic rays: A window on postinflationary reheating  epoch
  Phys.\ Atom.\ Nucl.\  {\bf 61}, 1028 (1998)
  [Yad.\ Fiz.\  {\bf 61}, 1122 (1998)]
  [arXiv:astro-ph/9709187].

\bibitem{Birkel:1998nx}
  M.~Birkel and S.~Sarkar,
  Astropart.\ Phys.\  {\bf 9}, 297 (1998)
  [arXiv:hep-ph/9804285].

\bibitem{Berezinsky:2000up}
  V.~Berezinsky and M.~Kachelriess,
  Phys.\ Rev.\ D {\bf 63}, 034007 (2001)
  [arXiv:hep-ph/0009053].

\bibitem{Barbot:2002ep}
  C.~Barbot and M.~Drees,
   ``Production of ultra-energetic cosmic rays through the decay of  super-heavy
  Phys.\ Lett.\ B {\bf 533}, 107 (2002)
  [arXiv:hep-ph/0202072].

\bibitem{Aloisio:2006yi}
  R.~Aloisio, V.~Berezinsky and M.~Kachelriess,
  Phys.\ Rev.\ D {\bf 74}, 023516 (2006)
  [arXiv:astro-ph/0604311].



\bibitem{GG}
H.~Georgi and S.~L.~Glashow, Phys.\ Rev.\ Lett.\ {\bf 32}, 438 (1974).

\bibitem{Barr:1981qv}
  S.~M.~Barr,
  Phys.\ Lett.\ B {\bf 112}, 219 (1982).
  
\bibitem{Derendinger:1983aj}
  J.~P.~Derendinger, J.~E.~Kim and D.~V.~Nanopoulos,
  Phys.\ Lett.\ B {\bf 139}, 170 (1984).
  
\bibitem{AEHN}
I.~Antoniadis, J.~R.~Ellis, J.~S.~Hagelin and D.~V.~Nanopoulos,
  Phys.\ Lett.\ B {\bf 231}, 65 (1989);
I.~Antoniadis, J.~R.~Ellis, J.~S.~Hagelin and D.~V.~Nanopoulos,
  Phys.\ Lett.\ B {\bf 208}, 209 (1988)
  [Addendum-ibid.\ B {\bf 213}, 562 (1988)];
I.~Antoniadis, J.~R.~Ellis, J.~S.~Hagelin and D.~V.~Nanopoulos,
  Phys.\ Lett.\ B {\bf 205}, 459 (1988);
I.~Antoniadis, J.~R.~Ellis, J.~S.~Hagelin and D.~V.~Nanopoulos,
  Phys.\ Lett.\ B {\bf 194}, 231 (1987);
I.~Antoniadis, J.~R.~Ellis, J.~S.~Hagelin and D.~V.~Nanopoulos,
  Phys.\ Lett.\ B {\bf 205}, 459 (1988).
  
\bibitem{Ellis:2002vk}
  J.~R.~Ellis, D.~V.~Nanopoulos and J.~Walker,
  Phys.\ Lett.\ B {\bf 550}, 99 (2002)
  [arXiv:hep-ph/0205336].
  
 \bibitem{Narain}
  K.~S.~Narain,
  Phys.\ Lett.\ B {\bf 169}, 41 (1986);
  K.~S.~Narain, M.~H.~Sarmadi and E.~Witten,
  Nucl.\ Phys.\ B {\bf 279}, 369 (1987).
  
\bibitem{Lopez:1995cs}
  J.~L.~Lopez and D.~V.~Nanopoulos,
  Phys.\ Rev.\ Lett.\  {\bf 76}, 1566 (1996)
  [arXiv:hep-ph/9511426].

\bibitem{Alon}
C.~Coriano, A.~E.~Faraggi and M.~Plumacher, Nucl.\ Phys.\ B {\bf 614}, 233 
(2001).

\bibitem{Sarkar:2001se}
  S.~Sarkar and R.~Toldra,
  Nucl.\ Phys.\ B {\bf 621}, 495 (2002)
  [arXiv:hep-ph/0108098].
  
\bibitem{Berezinsky:2005cq}
  V.~Berezinsky, A.~Z.~Gazizov and S.~I.~Grigorieva,
  Phys.\ Lett.\ B {\bf 612}, 147 (2005)
  [arXiv:astro-ph/0502550].
  
\bibitem{Ahlers}
  M.~Ahlers, L.~A.~Anchordoqui, H.~Goldberg, F.~Halzen, A.~Ringwald and
  T.~J.~Weiler,
  Phys.\ Rev.\ D {\bf 72}, 023001 (2005)
  [arXiv:astro-ph/0503229].

\bibitem{SarkarAPP}
  S.~Sarkar,
  Acta Phys.\ Polon.\ B {\bf 35}, 351 (2004)
  [arXiv:hep-ph/0312223].

\bibitem{Protheroe:1996si}
  R.~J.~Protheroe and P.~L.~Biermann,
  Astropart.\ Phys.\  {\bf 6}, 45 (1996)
  [Erratum-ibid.\  {\bf 7}, 181 (1997)]
  [arXiv:astro-ph/9605119].
    
\end{thebibliography}
\end{document}